\documentclass[pra,twocolumn,showpacs,nofootinbib,superscriptaddress,floatfix]{revtex4-1}
\usepackage{graphicx}
\usepackage{dcolumn}
\usepackage{bm}
\usepackage{comment}
\usepackage{amsmath,amscd,amssymb,color,amsfonts}
\usepackage{epstopdf}
\usepackage{float}
\usepackage{hyperref}
\usepackage{enumerate}

\usepackage{ulem}
\usepackage{xcolor}
\usepackage{mathtools}
\usepackage{afterpage}
\usepackage{ulem}
\usepackage{array}
\usepackage[T1]{fontenc}

\newcommand{\RNum}[1]{\uppercase\expandafter{\romannumeral #1\relax}}
\newcolumntype{P}[1]{>{\centering\arraybackslash}p{#1}}
\newcommand\T{\rule{0pt}{2.6ex}}       
\newcommand\B{\rule[-1.2ex]{0pt}{0pt}} 
\begin{document}
\title{ANN-Enhanced Detection of Multipartite Entanglement 
in a Three-Qubit NMR Quantum Processor} 
\author{Vaishali Gulati$^{\#}$}
\email{vaishali@iisermohali.ac.in}
\affiliation{Department of Physical Sciences, Indian
Institute of Science Education \& 
Research Mohali, Sector 81 SAS Nagar, 
Manauli PO 140306 Punjab India.}
\author{Shivanshu Siyanwal$^{\#}$}
\email{shivanshuacademics@gmail.com}
\affiliation{Department of Physical Sciences, Indian
Institute of Science Education \& 
Research Mohali, Sector 81 SAS Nagar, 
Manauli PO 140306 Punjab India.}
\author{Arvind}
\email{arvind@iisermohali.ac.in}
\affiliation{Department of Physical Sciences, Indian
Institute of Science Education \& 
Research Mohali, Sector 81 SAS Nagar, 
Manauli PO 140306 Punjab India.}
\author{Kavita Dorai}
\email{kavita@iisermohali.ac.in}
\affiliation{Department of Physical Sciences, Indian
Institute of Science Education \& 
Research Mohali, Sector 81 SAS Nagar, 
Manauli PO 140306 Punjab India.}
\begin{abstract}
We use an artificial neural network (ANN) model to identify the entanglement
class of an experimentally generated three-qubit pure state drawn from
one of the six inequivalent classes under stochastic local operations
and classical communication (SLOCC). The ANN model is also able to
detect the presence of genuinely multipartite entanglement (GME) in the
state.  We apply data science techniques to reduce the dimensionality
of the problem, which corresponds to a reduction in the number of
required density matrix elements to be computed.  The ANN model is
first trained on a simulated dataset containing randomly generated
states, and is later tested and validated on noisy experimental
three-qubit states cast in the canonical form and generated on a
nuclear magnetic resonance (NMR) quantum processor.  We benchmark the
ANN model via Support Vector Machines (SVMs) and K-Nearest Neighbor
(KNN) algorithms  and compare the results of our ANN-based entanglement
classification with existing three-qubit SLOCC entanglement
classification schemes such as 3-tangle  and correlation tensors.  Our
results demonstrate that the ANN model can perform GME detection and
SLOCC class identification with high accuracy, using {\em a priori}
knowledge of only a few density matrix elements as inputs.  Since the
ANN model works well with a reduced input dataset, it is an attractive
method for entanglement classification in real-life situations with
limited experimental data sets.  
\end{abstract}
\maketitle
\def\thefootnote{$\#$}\footnotetext{These authors
contributed equally to this
work}
\section{Introduction}
\label{sec1}
Quantum entanglement plays a pivotal role in various aspects
of quantum information processing such as quantum computing,
quantum cryptography, quantum metrology, and quantum
teleportation~\cite{horodecki-aps-2009}.  The most reliable
method for detecting entanglement is through full quantum
state tomography~\cite{nielsen-cam-2000}, however, this
approach is experimentally costly due to the exponential
increase in required projections with the dimension of the
corresponding Hilbert space~\cite{oliviera-els-2007}.  For
two-qubit and qubit-qutrit systems, the Peres-Horodecki
(PPT) criterion provides both necessary and sufficient
conditions for detecting
entanglement~\cite{horodecki-pla-1996}.  In the multipartite
case there is no simple necessary and sufficient condition
for separability\cite{doherty-prl-2002,doherty-pra-2004},
and  entanglement detection and characterization is
considered to be computationally
``NP-hard''~\cite{Gurvits-proceeding-2003,Gharibian-qic-2010}.

In recent years, there has been a growing interest in
experimental characterization of entanglement in various
physical systems such as
optics~\cite{hu-prl-2021,rivera-pra-2022}, trapped
ions~\cite{Huwer-njp-2013,Krutyanskiy-prl-2023}, cold
atoms~\cite{Bergh-pra-2021,Bergschneider-np-2019}, N-V
Centers~\cite{Rzepkowski-qip-2020} and superconducting
qubits~\cite{dany-science-2020}.  Entanglement protection
has emerged as a key area of research in this domain and has
been demonstrated in systems such as trapped
ions~\cite{Dumitrescu-nature-2022,Rodriguez-pra-2023},
superconducting qubits~\cite{yan-prl-2022} and
optics~\cite{Liao-jpb-2013}.  NMR is a versatile platform to
study entanglement and other non-local correlations in
nuclear spins using techniques such as local
measurements~\cite{amandeep-pla-2019,amandeep-pra-2016} and
expectation values of Pauli
operators\cite{amandeep-qip-2018,amandeep-pra-2018}.
Maximally entangled states such as
Greenberger-Horne-Zeilinger (GHZ) and W-type states have
been experimentally generated using
NMR~\cite{shruti-pra-2015,debmalya-pra-2015} and protected
from noise using dynamical decoupling
methods~\cite{harpreet-pra-2018,kondo-njp-2016}.

The integration of artificial intelligence and quantum
information processing has led to new breakthroughs in
solving resource-intensive problems~\cite{lohani-iop-2022}.
Machine learning and deep learning methods have been
employed for quantum state
tomography~\cite{torlai-spr-2018,dominik-pra-2022} and
entanglement classification and
detection~\cite{lu-aps-2018,roik-pla-2022}.  Other studies
have explored the utility of SVMs (Support Vector
Machines)~\cite{vintskevich-pra-2023}, Autoencoders
\cite{yosef-iop-2020} and GANs (Generative Adversarial
Networks)~\cite{chen-iop-2021} for quantum information
processing.  ANN-assisted quantum state tomography has been
shown to outperform standard tomographic methods in
high-dimensional photonic quantum
states~\cite{palmieri-npj-2022} as well in NMR
systems~\cite{akshay-arxiv-2023}. Various ANN architectures
have been deployed to study different aspects of
entanglement, for instance, to predict multipartite
entanglement structure~\cite{tian-aqt-2022}, to deduce the
entropy of highly entangled states~\cite{Harney-njp-2020},
to classify bound entangled states in a system of two
qutrits~\cite{goes-qip-2021}, and to generate artificial
entanglement witnesses for two and three qubits and for
entangled states in qudits~\cite{greenwood-pra-2023}.

In this study, we design and implement an artificial neural
network (ANN) model to detect and characterize entanglement
in three-qubit systems. We train the ANN on numerically
generated pure three-qubit quantum states to identify
genuine multipartite entanglement (GME) and classify the
states into one of six SLOCC classes: fully separable (SEP),
biseparable types 1, 2, and 3 (BS1, BS2, BS3), W, and GHZ
states. By representing states in their canonical form, we
reduce the number of density matrix elements from 128 to 18
essential ones, which are then used as input features for
the ANN model. We rank these features using ANOVA F-values
and then train multiple ANN models: the first model uses all
18 features, the next uses the top 17 features, and so on,
until the last model uses only the top feature. This allows
us to evaluate the performance of the models with varying
numbers of input features. We validate our approach by
preparing 30 three-qubit states on an NMR quantum processor,
splitting them into 18 for validation and 12 for testing. We
use 100 unique 'validation:test' combinations to tune and
test our 18 ANN models, with validation states used for
tuning and test states kept unknown to the ANN. We compare
our ANN results with traditional machine learning methods
(SVM, KNN)~\cite{cervantes-neuro-2020, hiesmayr-nsr-2021}
and other entanglement classification techniques,
demonstrating that ANN models with just 4 features can
detect GME and with 6 features can classify SLOCC for
unknown states efficiently. This study highlights the
effectiveness of ANN models in simplifying and accurately
classifying complex quantum states.

The rest of this paper is organized as follows:
Section~\ref{theory} provides a brief overview of the
theoretical background. Methods for generating three-qubit
generic states and the training dataset for ANN models are
discussed in Section~\ref{3qubit}, while SLOCC entanglement
classification and genuine entanglement detection using
correlation tensors and 3-tangle are covered in
Section~\ref{corr-tensor}, respectively.  Section~\ref{ann}
provides a detailed description of the ANN model designed
for classifying the entanglement class of three qubits. The
specific design of the ANN is outlined in
Section~\ref{ann-scheme}, while Sections~\ref{training} to
\ref{metric} cover the preparation of the training dataset,
dimensionality reduction, preparation of the experimental
dataset, and performance metrics of the ANN model,
respectively. Section~\ref{results} provides a detailed
description of the results. Section~\ref{univariate}
discusses the ANOVA F-value based univariate feature
selection method.  Section~\ref{classify} presents the
results of entanglement classification via ANN, and
Section~\ref{gme} contains the results of GME certification
via ANN.  Section~\ref{benchmark} compares the results of
our ANN model with other entanglement classifiers.
Section~\ref{svm} presents the comparison of our ANN model
with entanglement classification using SVM and KNN, while
Section~\ref{3tangle} contains the comparison with 3-tangle
and correlation tensors. Section~\ref{concl} offers some
concluding remarks. 
\section{Preliminary Background}
\label{theory}
\subsection{Generating Three-Qubit Training Datasets}
\label{3qubit}
We use the canonical form for three-qubit states to generate
random entangled
states~\cite{acin-phya-2001}: 
\begin{eqnarray}
&\!\!\!|\psi\rangle= \lambda_0|000\rangle+\lambda_1 e^{i \varphi}|100\rangle
+\lambda_2|101\rangle+\lambda_3|110\rangle+\lambda_4|111\rangle
\label{eq_acin} \!\!\!\!\!\nonumber \\
&{\displaystyle \sum_{i=0}^{i=4}}|\lambda_{i}|^{2} = 1,
\quad \lambda_{i}
\in \mathbb{R}, \quad \lambda_{i} \geq 0,\quad 0 \leq \varphi
\leq \pi.
\end{eqnarray}
 The canonical form
for three-qubit states 
can
be used to generate states from the six SLOCC inequivalent
entanglement classes~\cite{dur-pra-2000}.  To obtain the GHZ
class from the canonical form, the condition $\lambda_0
\lambda_4 \neq 0$ is implemented on the coefficients. The
different SLOCC inequivalent entanglement classes generated
from the canonical form~\cite{Li-qip-2018} are given in
Table~\ref{table_1}.

To generate the training dataset, we require randomly
sampled and normalized coefficients to ensure the generation
of states from all the SLOCC classes.  For this purpose, we
sampled the coefficients $\{\lambda_{i}\}_{i=0}^{i=4}$ from
a continuous uniform distribution of $10^{16}$ points
between the half-open interval denoted by $U_{(a,b]}$ with
$a = 0$ (a = 0 being excluded) and $b = 1$.  By strictly
keeping the relevant coefficients $\lambda_{i}$'s greater
than 0 for each SLOCC inequivalent entanglement class, we
prevent overlaps between the generated classes and also
avoid biases towards any specific basis within an
entanglement class~\cite{maziero-spr-2015}.  This approach
ensured a balanced representation of states across all SLOCC
inequivalent entanglement classes.  The required states were
generated using Numpy's random generator function to sample
the coefficients from the distribution, followed by
normalization.

\begin{table}
	\begin{tabular}{|l|l|p{1.5cm}|} 
\hline
\hline
~~$\boldsymbol{\lambda_{0}}$ and $\boldsymbol{\lambda_{4}}$ &
~~$\boldsymbol{\lambda_{1}}$, $\boldsymbol{\lambda_{2}}$,
$\boldsymbol{\lambda_{3}}$ and $\boldsymbol{e^{i\varphi}}$ &
\textbf{SLOCC Classes} 
\\ \hline \hline
$\lambda _{0}=0,\lambda _{4}\neq 0$ & ~~$\lambda _{2}\lambda
_{3}=\lambda_{1}\lambda _{4}e^{i\varphi }$ & 
A-B-C       \\
& ~~$\lambda _{2}\lambda _{3}\neq \lambda _{1}\lambda
_{4}e^{i\varphi }$ & 
A-BC        \\
\hline
$\lambda _{0}\neq 0,\lambda _{4}=0$ & ~~$\lambda
_{2}=\lambda _{3}=0$ &
A-B-C       \\
& ~~$\lambda _{2}=0,\lambda _{3}\neq 0$ &
C-AB       \\
& ~~$\lambda _{2}\neq 0,\lambda _{3}=0$ &
B-AC       \\
& ~~$\lambda _{2}\lambda _{3}\neq 0$ &
W          \\
\hline
$\lambda _{0}=\lambda _{4}=0$ & ~~$\lambda _{2}\lambda _{3}=0$ &
A-B-C      \\
& ~~$\lambda _{2}\lambda _{3}\neq 0$  &
A-BC       \\
\hline
$\lambda _{0}\lambda _{4}\neq 0$ & &
GHZ        \\
\hline
\end{tabular}
\caption{Coefficient parameterization for the generation of six SLOCC
inequivalent entanglement classes from the canonical
form~\cite{Li-qip-2018}.}
\label{table_1}
\end{table}
\subsection{SLOCC and GME Classification}
\label{corr-tensor} 
We used ranks of correlation tensors~\cite{gulati-epj-2022} 
and 3-tangle~\cite{dur-pra-2000} to
compare the results obtained from the ANN model. Since the
ranks of correlation tensors can distinguish between GME,
biseparable, and separable states, they are directly used
for comparing GME/non-GME classification results obtained
from ANN models. To compare SLOCC classification results
obtained from ANN, we use the 3-tangle in addition to the
ranks to distinguish between GHZ and W classes. 

Consider a three-qubit density matrix $\rho$ in the Hilbert
space ${\cal H}={\cal H}_1^2 \otimes {\cal H}_2^2 \otimes
{\cal H}_3^2$ where ${\cal H}^2$ denotes the two-dimensional
single-qubit Hilbert space. Let $\sigma_i,\, i=1,2,3$ denote
the generators of the unitary group SU(2), which together
with $\sigma_0=I$ ($I$ being a $2 \times 2$ identity
matrix), form an orthonormal basis of Hermitian operators.
Any density matrix $\rho$ can be decomposed
as~\cite{gulati-epj-2022}: \begin{eqnarray} && \rho =
\frac{1}{8}\left[ I \otimes I \otimes I \right.\nonumber \\
&&+ \sum t_{i}^{1} \sigma_i \otimes I \otimes I + \sum
t_{j}^{2} I \otimes \sigma_j \otimes I +  \sum t_{k}^{3} I
\otimes I \otimes \sigma_k \nonumber \\ &&+ \sum t_{ij}^{12}
\sigma_i \otimes \sigma_j \otimes I + \sum t_{ik}^{13}
\sigma_i \otimes I \otimes \sigma_k + \nonumber \\ && \sum
t_{jk}^{23} I \otimes \sigma_j \sigma_k  +  \sum
t_{ijk}^{123} \sigma_i \otimes \sigma_j \otimes \sigma_k ]
\label{decomposition} \end{eqnarray} $\rho$ can be
completely characterized by the expectation values:
$t_{i}^{1} = {\rm tr}(\rho \sigma_i \otimes I \otimes I)$,
$t_{j}^{2} = {\rm tr}(\rho I \otimes \sigma_j \otimes I)$,
$t_{k}^{3} = {\rm tr}(\rho I \otimes I \otimes \sigma_k)$,
$t_{ij}^{12} = {\rm tr}(\rho \sigma_i \otimes \sigma_j
\otimes I)$, $t_{ik}^{13} = {\rm tr}(\rho \sigma_i \otimes I
\otimes \sigma_k)$, $t_{jk}^{23} = {\rm tr}(\rho I \otimes
\sigma_j \otimes \sigma_k)$, $t_{ijk}^{123} = {\rm tr}(\rho
\sigma_i \otimes \sigma_j \otimes \sigma_k)$. The
expectation values $ t_{i}^{1}, t_{j}^{2}, t_{k}^{3} $ are
components of tensors of rank one denoted by $T^{(1)},
T^{(2)}, T^{(3)}$; $t_{ij}^{12}, t_{ik}^{13}, t_{jk}^{23}$
are components of tensors of rank two denoted by
$T^{(12)}, T^{(13)}, T^{(23)}$, and $t_{ijk}^{123}$ are
components of a rank three tensor $T^{123}$.  $T^{(qp)}$ are
two-qubit correlation tensors and $T^{(lmn)}$ is a
three-qubit correlation tensor. The correlation matrices are
computed from 13 expectation values obtained experimentally,
and the ranks of the computed matrices can be used to
classify a given state into 5 SLOCC inequivalent
entanglement classes.

The ranks of the correlation matrices and the corresponding
entanglement classes are given in Table~\ref{table_3}, which
can be used to classify the experimentally generated states
as GME/Non-GME~\cite{gulati-epj-2022}.  If the ranks of all
the correlation tensors are either 2 or 3, it indicates that
the state belongs to the GME class. 

The state exhibiting GME can be of two types: GHZ and W. The
measure of tripartite entanglement that can distinguish
between these two classes is the 3-tangle. We use the form
of the $3$-tangle based on the five-term canonical form for
tripartite systems: $\tau_{123} =
4|\lambda_0|^2|\lambda_4|^2 = 4 \langle 000|\rho|000 \rangle
\langle 111|\rho|111 \rangle)$~\cite{salinas-ent-2020} The
GHZ state has a non-zero 3-tangle. In contrast, the W state
has zero tangle. Moreover, 3-tangle calculations have
previously been used in the experimental classification of
entanglement in three-qubit NMR
states~\cite{amandeep-pra-2018}.

\begin{table}[h]
\begin{tabular}{cccl}
\hline \hline 
\textbf{R}$\boldsymbol{(T_{\underline{1}23})}~$ & 
\textbf{~R}$\boldsymbol{(T_{\underline{2}31})}~$ & 
\textbf{~R}$\boldsymbol{(T_{\underline{3}12})}~$ & 
\textbf{Class} \\
\hline 3 & 3 & 3 & Genuinely Entangled(GME)   \\
\hline 2 & 2 & 2 & Genuinely Entangled(GME)   \\
\hline 1 & 3 & 3 & Biseparable Type-1(BS${1}$)\\
\hline 3 & 1 & 3 & Biseparable Type-2(BS${2}$)\\
\hline 3 & 3 & 1 & Biseparable Type-3(BS${3}$)\\
\hline 1 & 1 & 1 & Fully Separable (SEP)      \\
\hline \hline 
\end{tabular}
\caption{Ranks (R[T$_{\underline{i}jk}$]) of correlation
matrices and the corresponding entanglement class of 
three-qubit pure states.} 
\label{table_3}
\end{table}
\begin{figure*}[h!]
\centering
\includegraphics[scale=1.0]{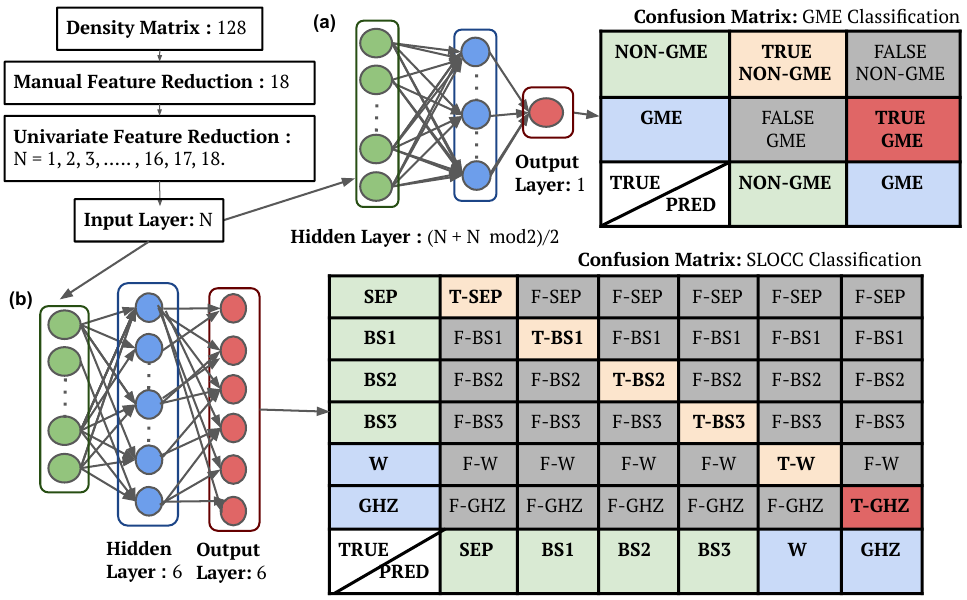}
\caption{(Color online) The ANN-based entanglement
classification protocol. The density matrix $\rho$ with 128
features  (i.e, 64 real entries and 64 imaginary entries) is
processed via manual and univariate feature selection to
generate a $N \times 1$ dimensional input vector with $N =
1,2,3, \dots 17, 18$. (a) The GME ANN model consists of $N$
neuron input layers ($L_{1}$ in green), $(N + N \,{\rm
mod}\, 2)/2$ neuron hidden layers ($L_{2}$ in blue) and $1$
neuron output layer ($L_{3}$ in red). (b) The SLOCC ANN
model consists of $N$ neuron input layers ($L_{1}$ in
green), $6$ neuron hidden layers ($L_{2}$ in blue) and $6$
neuron output layers ($L_{3}$ in red).  } 
\label{ANN-figure}
\end{figure*} 
\section{ANN Model for Entanglement Classification} 
\label{ann} 
\subsection{Basic ANN Architecture} 
\label{ann-scheme} 
A basic feed-forward
artificial neural network (FFNN) consists of an input layer
(responsible for loading the data features), hidden layers
(which learn the weights and biases of the network), and an
output layer (which predicts the labels corresponding to the
data features). 
The output from a neuron is determined by the activation 
function, which could for example be  a ReLU, linear or
a sigmoid~\cite{bern-cam-2022}.
To avoid overfitting, we have
restricted ourselves to a smaller number of features via
univariate feature selection and small ANN models.
Overfitting is characterized by increasing validation loss
and decreasing training loss, where the loss represents the
error between predicted and true labels for the validation
and training data sets. The neural network optimizes the
weights and biases of its underlying function (a process
known as `learning') by receiving feedback about incorrectly
and correctly predicted labels via back propagation, which
results in the minimization of loss or cost functions over
the validation and training set as the training progresses
through the epochs~\cite{bern-cam-2022}.

A schematic of the ANN model architecture is given in
Figure~\ref{ANN-figure}.  The density matrix $\rho$ is first
separated into its real and imaginary components, followed
by flattening of the two $8 \times 8$ matrices into a single
$128 \times 1$ vector.  The state is then written in the
canonical form in order to generate an $18 \times 1$ input
vector, followed by feature reduction via ANOVA, resulting
in a decreased dimensionality (features) of the problem to
$N \leq 18$.  These transformed input vectors are then fed
into the SLOCC and GME ANN entanglement classifiers,
yielding probabilities of a random state as belonging to one
of these entanglement classes.  For the GME case, the green
circles in Figure~\ref{ANN-figure} denote the input layer of
$N$ neurons with $N = 1,2,3,..,17, 18$, the blue circles
denote the hidden layer of $(N + N\, {\rm mod}\, 2)/2$ neurons,
and the red circles denote the final output layer of 1
neuron  which outputs with a probability $ 0 \leq p_{0} \leq
1$ whether the input state belongs to the GME class or not. 

For the binary classification problem, we set the number of
neurons in the hidden layer to be approximately half of the
input layer, following the formula $(N + N\,{\rm mod}\, 2)/2$.
This approach aligns with the rule of thumb in applied
machine learning that the number of neurons in the hidden
layer should be between the input feature count and the
output class count~\cite{gaurang-ijcte-2011}.

States with $p_0 < 0.5$ are considered Non-GME (labeled `0')
while states with $p_0 \geq 0.5$ are considered GME (labeled
`1').  The results are represented in the confusion matrix
in Figure~\ref{ANN-figure}(a), which gives the
classification errors (given by ``FALSE NON-GME'' and
``FALSE GME''). 

Figure~\ref{ANN-figure}(b) describes the ANN architecture
and protocol for the classification of the input density
matrix $\rho$ into the six inequivalent SLOCC entanglement
classes.  The SLOCC ANN architecture is initialized with an
input layer of $N$ neurons followed by a hidden layer with 6
neurons, which we found to be optimal through trial and
error, following the guiding principles mentioned
in~\cite{gaurang-ijcte-2011}.  The final output layer
consists of $6$ neurons, each representing the probability
set $\{p_{1}, p_{2}, p_{3}, p_{4}, p_{5}, p_{6}\}$
corresponding to the set $\{$ SEP, BS1, BS2, BS3, W, GHZ
$\}$  where $\sum_{i}^{6} p_{i} = 1 $, the maximum of the
set $\{p_{i}\}_{i=1}^{6}$ is considered for assigning the
prediction labels. For a state in the test set, if the class
label associated with the maximum probability (i.e., label
predicted by the ANN) corresponds to the true class label of
the state, then it is considered correctly classified (i.e.,
ANN accuracy = $1$) otherwise it is considered incorrectly
classified (ANN accuracy = $0$). The predicted labels by ANN
and true class labels are compared in the extended confusion
matrix given in Figure~\ref{ANN-figure}(b), where, ``T-SEP''
and ``F-SEP'' correspond to a correctly classified SEP state
and a mis-classified SEP state, respectively.  The same
logic holds for the other entanglement classes.  For GME
classification, the neural network topology is $L_{1} \times
L_{2} \times L_{3} = N \times {\displaystyle \frac{N + 
N \,{\rm mod}\,\, 2}{2}} \times 1$,  
referring to the input, hidden, and output
layers, respectively. For SLOCC inequivalent entanglement
class categorization, the topology takes the form  of $N
\times 6 \times 6$ with an input layer of $N$ neurons
corresponding to the total number of features.  

\subsection{Labeling the Training Dataset} 
\label{training} 
After numerical generation of random three-qubit states from
the six inequivalent SLOCC classes, the next step is to
convert the labels of the generated states into a form that
the neural network can process. The generated states are
assigned labels according to the type of entanglement
classification problem.  Let $\rho_{}$ denote the density
matrix belonging to one of the SLOCC inequivalent
entanglement classes.  For the SLOCC inequivalent
entanglement classification problem, we encode the class
labels via one-hot-encoding which is done via assigning $1$
or $0$ at the position corresponding to the SLOCC label in a
six-element row vector $\vec{H}(\rho)$ initialized with
zeros (the first position is for SEP, second for BS$_{1}$,
third for BS$_{2}$, and so on), where, $\vec{H}(\rho_{})$ =
[$H_{i}(\rho)$], $H_{i}(\rho)$ $\in  \left \{ 0,1\right \}$
and $\sum_{i=1}^{6}H_{i}(\rho)=1$.  For the GME/NON GME
classification, we encode the GME label via binary encoding
in a single element vector $\vec{B}(\rho)$ (1 is for GME
while 0 is for Non-GME). In order to do a comparative
analysis with SVM and KNN algorithms, integer encoding of
the SLOCC labels is required, which is done by assigning the
values from the set $\{0,1,2,3,4,5\}$ to the single element
vector $\vec{I}(\rho)$($0$ for SEP, $1$ for BS$_1$, ..., $5$
for GHZ). The complete label row vector is given by
$\vec{L_{\rho}} =
[\vec{H(\rho_{}}))~~\vec{B}(\rho)~~\vec{I}(\rho)]$.  The
density matrix $\rho$ is then decomposed into real and
imaginary component matrices followed by flattening,
resulting in two $64 \times 1$ row vectors
$\vec{V}_{\mathrm{Re}}(\rho)$ and
$\vec{V}_{\mathrm{Im}}(\rho)$ containing the
${\mathrm{Re}}(\rho)$ and ${\mathrm{Im}}(\rho)$ components,
structured as: 

\begin{equation}
\vec{V}_{\mathrm{Re}}(\rho)=\left[\operatorname{Re}\left(\rho_{00}\right)
\operatorname{Re}\left(\rho_{01}\right) \ldots . \operatorname{Re}\left(\rho_{i
J}\right) \ldots . \operatorname{Re}\left(\rho_{77}\right)\right] 
\end{equation}

\begin{equation}
\vec{V}_{\mathrm{Im}}(\rho)=\left[\operatorname{Im}\left(\rho_{00}\right)
\operatorname{Im}\left(\rho_{01}\right) \ldots . \operatorname{Im}\left(\rho_{i
J}\right) \ldots . \operatorname{Im}\left(\rho_{77}\right)\right] 
\end{equation}

Along with the encoded labels, the data vector for the 
state $\rho$ is given by:
\begin{equation}
\label{eq-data}
\vec{D}_{\rho} = [\vec{V}_{\mathrm{Re}}(\rho)
~~\vec{V_{\mathrm{Im}}}(\rho)~~\vec{H}(\rho_{})~~\vec{B}(\rho)~~\vec{I}(\rho)]
\end{equation} 
\subsection{Dimensionality Reduction} 
\label{dim-reduce}

To reduce the dimensionality of the entanglement classification problems,
redundant features were removed by identifying irrelevant density matrix
elements. Assuming a linear relationship between the input features (i.e.,
density matrix elements) and the learned ANN function $f(\rho)$, the
relationship can be expressed as:
\begin{equation}
f(\rho) = \sum_{i=0}^{7}\sum_{j=0}^{7}(\alpha_{ij}R_{ij} + \beta_{ij}I_{ij}) +
\gamma
\label{eq8}
\end{equation}
where $\{\alpha_{ij}, \beta_{ij}\}$ are the set of 128 coefficients which the
ANN learns about and optimizes during the training procedure, and $\gamma$ is a
constant bias term. Since, canonical form for three-qubit states
(Eq.~\ref{eq_acin}) consists of only 5 non-zero terms, we are left with 18 (4
imaginary and 14 real) density matrix elements. The purpose of reduction is to
simplify the problem and improve the model's performance as well as to prevent
the ANN model from learning trivial patterns which are already present in the
dataset.  We further apply univariate selection over these remaining 18
features and assign scores to each of the features, which are then fed as input
to the ANN model.
\subsection{Preparing the Experimental Dataset}
\label{expt}
The three ${}^{19}\mathrm{F}$ nuclei in the molecule trifluoroiodoethylene were
used to experimentally realize three qubits, with relaxation times in the range
$1 s\leq\langle T_{1}\rangle \leq 7s $.  The experiments were performed on a
Bruker AVANCE-III $400 \mathrm{MHz}$ NMR spectrometer equipped with a BBO probe at
temperature $T \approx 298 \mathrm{~K}$.  In the high-temperature and high-field
approximation, assuming a weak scalar coupling  $J_{ij}$ between the $i$th and
$j$th spins, the Hamiltonian for the three-qubit NMR system is given
by~\cite{oliviera-els-2007}: 
\begin{align} 
\mathcal{H}=-\sum_{i=1}^{3}
\omega_{i} I_{i z}+2 \pi \sum_{i<j}^{3} J_{i j} I_{i z} I_{j z} 
\end{align}
where $\omega_{i}$  refers to the chemical shift of the $i$th spin with the
experimentally determined scalar couplings being $\mathrm{J}_{23}=-128.32
\mathrm{~Hz}$, $\mathrm{~J}_{13}=47.67 \mathrm{~Hz}$ and $\mathrm{J}_{12}=69.65
\mathrm{~Hz}$. 
The spatial averaging technique was used to initialize the system into 
a pseudopure (PPS) state~\cite{mitra-magn-2007,cory-els-1998}: 
\begin{align}
\rho_{000}=\frac{(1-\epsilon)}{8} \mathbb{I}_{8}+\epsilon|000\rangle\langle 000|
\end{align}
where  $\mathbb{I}_{8}$ is the $8 \times 8$ identity operator and $\epsilon \sim
10^{-5}$ is the spin polarization at $T \approx 298 \mathrm{~K}$.

\begin{figure*}
\centering
\includegraphics[scale=1]{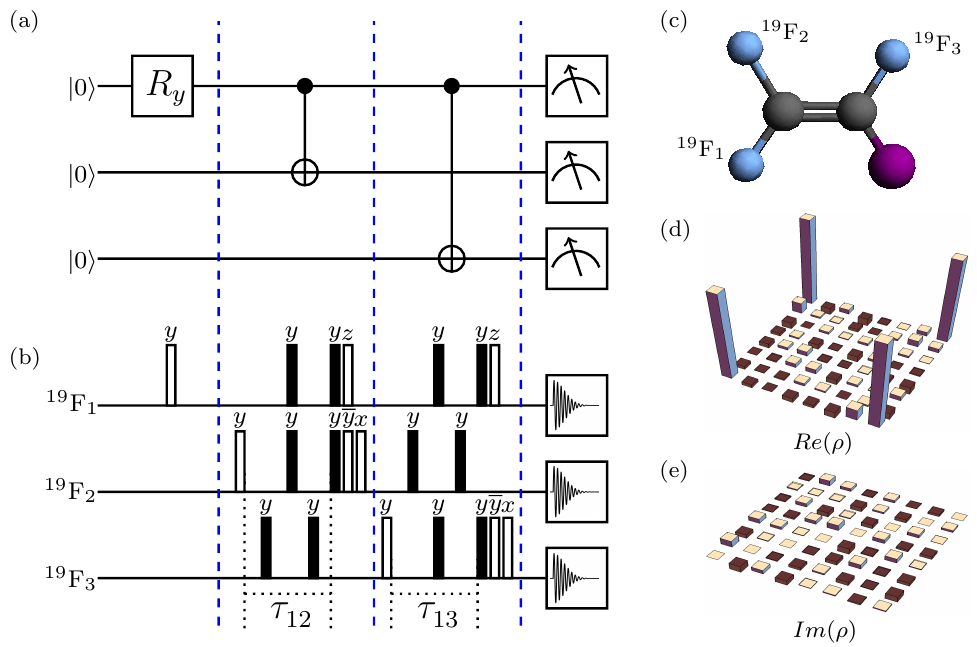}
\caption{(Color online) Generation of 
the experimental data: (a) Quantum circuit used to prepare a
three-qubit state, in this case, 
the GHZ state. (b) The corresponding NMR pulse sequence,
with rectangles representing 
rf pulses of different phases and angles. The phases of each pulse are written
above each pulse, with filled rectangles representing $\pi/2$ pulses and
unfilled rectangles representing $\pi$ pulses;
the time period ${\displaystyle \tau=\frac{1}{2J}}$.
(c) The 
trifluoroiodoethylene molecule with 
three NMR active $^{19}F$ nuclei is used as the
three-qubit system. (d),(e) Bar plots of the real and imaginary
parts of the experimentally obtained density matrix.}
\label{Exp_fig}
\end{figure*}

Random three-qubit states were theoretically generated for each SLOCC class
using the Mathematica package~\cite{Mathematica}, and the gate sequences of
these random states were prepared via the open source Mathematica package
UniversalQCompiler~\cite{iten-arxiv}.  For the experimental implementation of
these states, we generated unitaries with the help of the GRAPE
package~\cite{Khaneja-jmr-2005}.  
The experimental
protocol is schematically depicted in Figure~\ref{Exp_fig}, for an illustrative
GHZ state.  A radio frequency (rf) pulse sequence of varying angles and phases
combined with $J$-evolution periods was used to implement single and two-qubit
gates, followed by the measurement step. 
Constrained convex optimization (CCO)
tomography was used to reconstruct the
final experimental density matrix~\cite{akshay-qip-2021} protocol. The bar plots of real and imaginary
components of the experimental density matrix of the GHZ state are shown in
Figures~\ref{Exp_fig}(d) and (e), respectively.  

The fidelity measure~\cite{zhang-prl-2014} used to
compute the state fidelities of the experimentally 
reconstructed states is given by:
\begin{align}
\mathcal{F}\left(\rho_{\text {expt }}, \rho_{\text {theo
}}\right)=\frac{\left|\operatorname{Tr}\left[\rho_{\text {expt }} \rho_{\text
{theo }}^{\dagger}\right]\right|}{\sqrt{\operatorname{Tr}\left[\rho_{\text {expt
}}^{\dagger} \rho_{\text {expt }}\right] \operatorname{Tr}\left[\rho_{\text
{theo }}^{\dagger} \rho_{\text {theo }}\right]}}
\end{align}
where  $\rho_{\text{expt}}$ and $\rho_{\text {theo }}$  are the experimentally
obtained and the theoretically computed density matrices, respectively.  We
obtained good experimental fidelities in the range ($0.87-0.98$) for all the
states. 

\subsection{ANN Performance Metrics}
\label{metric} 
We used the metric Accuracy ($A$) 
to evaluate the performance of the ANN, SVM and
KNN models given by:
\begin{equation}
A  = \frac{(TP+TN)}{(TP+TN+FP+FN)}
\label{eq-metric-accuracy}
\end{equation} 
where $TP$ represent True Positives, $TN$ represent True Negatives, $FP$
represent False Positives, and $FN$ represent False Negatives.
In Figure~\ref{ANN-figure}, ``TRUE GME'' corresponds to $TP$,
``TRUE NON-GME'' to $TN$, ``FALSE GME'' to $FP$ 
and ``FALSE NON-GME'' to $FN$.
Alternatively, the accuracy $A$ can be calculated by 
summing the main diagonal entries and dividing by the total 
entries~\cite{diego-ai-2022,bern-cam-2022}. 

The NMR experiment dataset consists of 30 states with 5 states from each
SLOCC class (i.e., 10 states from GME and 20 from Non-GME classes).
To facilitate model optimization, we split this dataset
into two distinct sets: a validation set and a test set. The validation set
comprises 18 states, with 3 states randomly selected from each SLOCC class. This
separation is imperative as it allows us to fine-tune the 
hyperparameters of the
ANN model on a separate dataset, 
before applying it to the test set. Doing
so allows us to assess and enhance the model's performance without introducing
bias or overfitting. Furthermore, we use 12 states (the remaining 2 from
each class after construction of the validation set), exclusively for testing
purposes, keeping them concealed from the ANN model
during each training cycle. There are  
$({}^{5}C_{3})^{6} = 10^{6}$ possible unique
validation and test set combinations, from 
which we randomly sampled 100 combinations (which can be considered as
100 separate training instances). We train, validate and test all the ANN models
from $N=1$ to $N=18$ over these 100 combinations and arrive at mean accuracy
$\mu(A)$ and standard error $\sigma(A)$. We project the overall accuracy and
error over entire 30 states via the following:
\begin{equation}
\label{eq-metric-11}
\mu(A) = 0.6\mu(A_{v}) + 0.4\mu(A_{t})  
\end{equation}
\begin{equation}\label{eq-metric-12}
\sigma(A)^{2} = 0.36\sigma(A_{v})^{2} 
+ 0.16\sigma(A_{t})^{2} + 0.48\sigma(A_{v}A_{t})
\end{equation} 
where $(\mu(A_{v}),\sigma(A_{v}))$, $(\mu(A_{t}),\sigma(A_{t}))$ refer to the
average accuracy and standard error over 
the validation and test sets, respectively.
We used the error propagation formula for $F = aX + bY$ with 
$\sigma(F) =
\left[ a^{2}\sigma({X})^{2} + b^{2}\sigma({Y})^{2} +
2ab\,\sigma({XY})\right]^{-\frac{1}{2}}$ with $a =
{18}/{30} = 0.6$, $b = {12}/{30} = 0.4$ and $\sigma({XY})
=
\delta({XY})\sigma({X})\sigma({Y})$ as the covariance and
$\delta({XY})$ as the
correlation between variables $X$ and $Y$. 
\section{Results and Discussion} 
\label{results}
\subsection{Univariate Selection}
\label{univariate}
Univariate selection is a feature selection technique used in machine learning
to evaluate and select features based on their individual relationship with the
target variable. It involves statistical tests to determine the significance of
each feature independently, without considering the interactions between
features. For continuous features and categorical targets, we commonly use the
ANOVA-F test
\cite{betty-epm-1974,gelman-as-2005,ostertagova-ajme-2013}.
After
computing the test statistics for all features, the features are ranked based
on their scores. It helps in simplifying the model and reducing the
computational load by narrowing down the feature set to the
most relevant ones.  The ANOVA-F value is calculated for
each feature by conducting an ANOVA test.  This test
assesses whether there are significant differences between
the means of the target variable for different levels of the
feature. Essentially, it measures the variance explained by
the feature in relation to the target variable.

\noindent Between-Group Variance: Variance of the target
variable between different groups formed by the feature.  

\noindent Within-Group Variance: Variance of the target
variable within each group formed by the feature.  The
F-value is the ratio of these two variances:
$$F=\frac{\displaystyle \text{variance between group means}}
{\displaystyle \text{variance within the groups }}$$

In our case, the features are the $18$ density matrix
elements that we obtained
earlier via dimensionality reduction and the target
variables are the GME and
SLOCC classes, $\vec{\boldsymbol{\mathrm{B(\rho)}}}$ and
$\vec{\boldsymbol{\mathrm{I(\rho)}}}$, respectively. We then
use ANOVA F-value
to assign scores to each of the
features~\cite{ostertagova-ajme-2013}. Since it
is a data driven technique, we numerically generated $1200$
states (with each
class represented by 200 states) for both the GME and the
SLOCC classification
problems via the method mentioned in Sec.~\ref{3qubit}, using ``f classif'' and
``Select K Best'' from the Sklearn library.  For GME and
SLOCC classification,
all the ANOVA F-values (feature scores) and corresponding
features are given in
Table~\ref{table_score_1}. We test whether ANN models with
fewer than 18
features can achieve accuracies comparable to the
full-feature models (with $N
= 18$) for both GME detection and SLOCC classification.  For
each of the
classification problems i.e., GME and SLOCC, the features
are rearranged in
decreasing order of their scores. For SLOCC, the feature
order is :
($\mathbf{Re}(\rho_{07})$, $ \mathbf{Re}(\rho_{06}) $, $
\mathbf{Re}(\rho_{05})
$, $ \mathbf{Re}(\rho_{67}) $, $ \mathbf{Re}(\rho_{57}) $, $
\mathbf{Re}(\rho_{56}) $, $ \mathbf{Re}(\rho_{55}) $, $
\mathbf{Re}(\rho_{00})
$, $ \mathbf{Re}(\rho_{66}) $, $ \mathbf{Re}(\rho_{44}) $, $
\mathbf{Im}(\rho_{46}) $, $ \mathbf{Im}(\rho_{04}) $, $
\mathbf{Im}(\rho_{45})
$, $ \mathbf{Re}(\rho_{46}) $, $ \mathbf{Re}(\rho_{04}) $, $
\mathbf{Re}(\rho_{45}) $, $ \mathbf{Re}(\rho_{47}) $, $
\mathbf{Im}(\rho_{47})
$).  For GME, the feature order is : [$
\mathbf{Re}(\rho_{07}) $, $
\mathbf{Re}(\rho_{56}) $, $ \mathbf{Re}(\rho_{05}) $, $
\mathbf{Re}(\rho_{06})
$, $ \mathbf{Re}(\rho_{57}) $, $ \mathbf{Re}(\rho_{67}) $, $
\mathbf{Re}(\rho_{44}) $, $ \mathbf{Re}(\rho_{46}) $, $
\mathbf{Re}(\rho_{00})
$, $ \mathbf{Im}(\rho_{47}) $, $ \mathbf{Re}(\rho_{45}) $, $
\mathbf{Re}(\rho_{04}) $, $ \mathbf{Im}(\rho_{04}) $, $
\mathbf{Im}(\rho_{45})
$, $ \mathbf{Re}(\rho_{47}) $, $ \mathbf{Im}(\rho_{46}) $, $
\mathbf{Re}(\rho_{66}) $, $ \mathbf{Re}(\rho_{55}) $].
These features are fed
into the input layer of the ANN models in the same order.
For example, the GME
ANN model $ 1 \times 1 \times 1 $ consists of 1 input feature
$\mathbf{Re}(\rho_{07})$, the model $ 2 \times 1 \times 1 $ consists of 2 input
features $\mathbf{Re}(\rho_{07})$ and $\mathbf{Re}(\rho_{56})$ and so on. The
same rule is followed when calculating SVM and KNN accuracies. 

\begin{table}
\begin{ruledtabular}
\begin{tabular}{ccrr}
Feature No. &Feature&GME&SLOCC\\
\hline
1 & $ \mathbf{Re}(\rho_{00}) $ & 0.08 & 70.43    \\
2 & $ \mathbf{Re}(\rho_{04}) $ & 0.19 & 1.27     \\
3 & $ \mathbf{Re}(\rho_{05}) $ & 208.61 & 291.39 \\
4 & $ \mathbf{Re}(\rho_{06}) $ & 195.36 & 283.19 \\
5 & $ \mathbf{Re}(\rho_{07}) $ & 391.70 & 315.85 \\
6 & $ \mathbf{Re}(\rho_{44}) $ & 53.90 & 25.85   \\
7 & $ \mathbf{Re}(\rho_{45}) $ & 1.20 & 0.40     \\
8 & $ \mathbf{Re}(\rho_{46}) $ & 0.20 & 1.48     \\
9 & $ \mathbf{Re}(\rho_{47}) $ & 2.31 & 2.54     \\
10 & $ \mathbf{Re}(\rho_{55}) $ & 0.12 & 80.63   \\
11 & $ \mathbf{Re}(\rho_{56}) $ & 286.44 & 110.98 \\
12 & $ \mathbf{Re}(\rho_{57}) $ & 92.00 & 98.17 \\
13 & $ \mathbf{Re}(\rho_{66}) $ & 11.17 & 58.51 \\
14 & $ \mathbf{Re}(\rho_{67}) $ & 44.76 & 95.44 \\
15 & $ \mathbf{Im}(\rho_{04}) $ & 0.24 & 0.38 \\
16 & $ \mathbf{Im}(\rho_{45}) $ & 0.00 & 1.49 \\
17 & $ \mathbf{Im}(\rho_{46}) $ & 0.03 & 0.25 \\
18 & $ \mathbf{Im}(\rho_{47}) $ & 2.30 & 0.34 \\
\end{tabular}
\end{ruledtabular}
\caption{ANOVA F-values (Feature Scores) for each
of the individual 18 features for GME detection and SLOCC classification.}
\label{table_score_1}
\end{table}

\begin{table*}
\begin{ruledtabular}
\begin{tabular}{rcrrrrrrrrr}
N & ($L_{1} \times L_{2} \times L_{3}$) & $P_{2}$ & $P_{3}$ & $\mu(A_{v}) \pm \sigma(A_{v}) $ & $\mu(L_{v}) \pm \sigma(L_{v}) $ & $\mu(A_{t}) \pm \sigma(A_{t})$ & $\mu(L_{t}) \pm \sigma(L_{t})$ & $A_{T} \pm \sigma(A_{T})$ & $A_{1}$ & $A_{2}$ \\ 
\hline
1 & $1	\times 6	\times 6$ & 7 & 12 & 0.431±0.080 & 1.600±0.126 & 0.330±0.105 & 1.428±0.013 & 0.391±0.099 & 0.333 & 0.233 \\
2 & $2	\times 6	\times 6$ & 8 & 12 & 0.629±0.090 & 1.276±0.207 & 0.560±0.129 & 1.063±0.090 & 0.601±0.106 & 0.600 & 0.400 \\
3 & $3	\times 6	\times 6$ & 9 & 12 & 0.809±0.059 & 1.250±0.217 & 0.782±0.105 & 0.905±0.202 & 0.798±0.094 & 0.767 & 0.633 \\
4 & $4	\times 6	\times 6$ & 10 & 12 & 0.794±0.052 & 1.285±0.210 & 0.792±0.076 & 0.893$\pm$0.199 & 0.794$\pm$0.087 & 0.767 & 0.667 \\
5 & $5	\times 6	\times 6$ & 11 & 12 & 0.795±0.055 & 1.201±0.203 & 0.776±0.067 & 0.890$\pm$0.185 & 0.787±0.087 & 0.767 & 0.667 \\
6 & $6	\times 6	\times 6$ & 12 & 12 & 0.876±0.057 & 1.114±0.220 & 0.854±0.088 & 0.822±0.200 & 0.867±0.090 & 0.833 & 0.733 \\
7 & $7	\times 6	\times 6$ & 13 & 12 & 0.798±0.063 & 1.061±0.211 & 0.783±0.087 & 0.791±0.154 & 0.792±0.092 & 0.800 & 0.833 \\
8 & $8	\times 6	\times 6$ & 14 & 12 & 0.815±0.072 & 1.058±0.172 & 0.793±0.090 & 0.890±0.206 & 0.806±0.094 & 0.800 & 0.800 \\
9 & $9	\times 6	\times 6$ & 15 & 12 & 0.826±0.053 & 1.048±0.153 & 0.802±0.089 & 0.898±0.223 & 0.816±0.089 & 0.833 & 0.833 \\
10 & $10	\times 6	\times 6$ & 16 & 12 & 0.812±0.063 & 1.094±0.165 & 0.792±0.089 & 0.853±0.209 & 0.804±0.092 & 0.833 & 0.767 \\
11 & $11	\times 6	\times 6$ & 17 & 12 & 0.818±0.060 & 1.102±0.180 & 0.789±0.094 & 0.869±0.216 & 0.806±0.092 & 0.833 & 0.767 \\
12 & $12	\times 6	\times 6$ & 18 & 12 & 0.819±0.057 & 1.088±0.165 & 0.788±0.092 & 0.859±0.208 & 0.807±0.091 & 0.833 & 0.767 \\
13 & $13	\times 6	\times 6$ & 19 & 12 & 0.818±0.061 & 1.096±0.169 & 0.787±0.091 & 0.859±0.208 & 0.806±0.092 & 0.833 & 0.767 \\
14 & $14	\times 6	\times 6$ & 20 & 12 & 0.814±0.058 & 1.103±0.175 & 0.788±0.092 & 0.852±0.209 & 0.803±0.091 & 0.833 & 0.767 \\
15 & $15	\times 6	\times 6$ & 21 & 12 & 0.811±0.065 & 1.102±0.195 & 0.792±0.088 & 0.855±0.201 & 0.803±0.092 & 0.833 & 0.767 \\
16 & $16	\times 6	\times 6$ & 22 & 12 & 0.813±0.063 & 1.109±0.163 & 0.786±0.090 & 0.857±0.212 & 0.802±0.092 & 0.833 & 0.767 \\
17 & $17	\times 6	\times 6$ & 23 & 12 & 0.812±0.063 & 1.112±0.179 & 0.788±0.089 & 0.855±0.205 & 0.803±0.092 & 0.833 & 0.800 \\
18 & $18	\times 6	\times 6$ & 24 & 12 & 0.815±0.059 & 1.107±0.161 & 0.789±0.091 & 0.850±0.202 & 0.805±0.091 & 0.833 & 0.800 \\
\end{tabular}
\end{ruledtabular}
\caption{SLOCC classification via ANN accuracy values for
total features from $N=1$ to $N=18$. The ANN network for
each $N$ value given by $N \times 6 \times 6$, $P_{2}$ and
$P_{3}$ refer to total learning parameters (i.e., weights
and biases) for the hidden and the output
layer,respectively.  $\mu(A_{v}) \pm \sigma(A_{v}) $ refers
to average validation set accuracy, $\mu(L_{v}) \pm
\sigma(L_{v}) $ to average validation set loss, $\mu(A_{t})
\pm \sigma(A_{t})$ refers to average test set accuracy and
$\mu(L_{t}) \pm \sigma(L_{t})$ to average test set loss,
with standard error values calculated over 100
validation-test set combinations.  $A_{T} \pm \sigma(A_{T})$
refers to overall accuracy over all the experimental states.
$A_{1}$ and $A_{2}$ refers to accuracies obtained via SVM
and KNN, respectively.
}
\label{slocc_ann_table}
\end{table*}

\begin{table*}
\begin{ruledtabular}
\begin{tabular}{rrrrrrrrrrr}
N & ($L_{1} \times L_{2} \times L_{3}$) & $P_{2}$ & $P_{3}$ & $\mu(A_{v}) \pm \sigma(A_{v}) $ & $\mu(L_{v}) \pm \sigma(L_{v}) $ & $\mu(A_{t}) \pm \sigma(A_{t})$ & $\mu(L_{t}) \pm \sigma(L_{t})$ & $A_{T} \pm \sigma(A_{T})$ & $A_{1}$ & $A_{2}$  \\
\hline
1 & $1	\times 1	\times 1$ & 2 & 2 & 0.826±0.027 & 0.62±0.06 & 0.817±0.048 & 0.525±0.035 & 0.823±0.038 & 0.833 & 0.400 \\
2 & $2	\times 1	\times 1$ & 3 & 2 & 0.878±0.072 & 0.501±0.125 & 0.862±0.078 & 0.4±0.063 & 0.872±0.061 & 0.900 & 0.567 \\
3 & $3	\times 2	\times 1$ & 5 & 3 & 0.878±0.055 & 0.415±0.096 & 0.847±0.072 & 0.305±0.045 & 0.866±0.052 & 0.933 & 0.833 \\
4 & $4	\times 2	\times 1$ & 6 & 3 & 0.959±0.046 & 0.374±0.053 & 0.948±0.057 & 0.317±0.046 & 0.954±0.046 & 0.967 & 0.933 \\
5 & $5	\times 3	\times 1$ & 8 & 4 & 0.964±0.037 & 0.368±0.042 & 0.95±0.057 & 0.31±0.043 & 0.958±0.043 & 0.933 & 0.933 \\
6 & $6	\times 3	\times 1$ & 9 & 4 & 0.881±0.058 & 0.421±0.086 & 0.865±0.074 & 0.328±0.06 & 0.874±0.054 & 0.733 & 1.000 \\
7 & $7	\times 4	\times 1$ & 11 & 5 & 0.872±0.058 & 0.383±0.102 & 0.866±0.082 & 0.304±0.062 & 0.869±0.056 & 0.733 & 1.000 \\
8 & $8	\times 4	\times 1$ & 12 & 5 & 0.876±0.06 & 0.386±0.104 & 0.865±0.081 & 0.298±0.056 & 0.872±0.056 & 0.733 & 1.000 \\
9 & $9	\times 5	\times 1$ & 14 & 6 & 0.892±0.06 & 0.351±0.103 & 0.879±0.076 & 0.281±0.06 & 0.887±0.055 & 0.733 & 0.967 \\
10 & $10	\times 5	\times 1$ & 15 & 6 & 0.891±0.062 & 0.348±0.095 & 0.892±0.073 & 0.275±0.055 & 0.891±0.055 & 0.733 & 0.967 \\
11 & $11	\times 6	\times 1$ & 17 & 7 & 0.916±0.049 & 0.33±0.066 & 0.899±0.069 & 0.287±0.065 & 0.909±0.049 & 0.733 & 1.000 \\
12 & $12	\times 6	\times 1$ & 18 & 7 & 0.919±0.041 & 0.326±0.054 & 0.889±0.071 & 0.286±0.066 & 0.907±0.047 & 0.733 & 1.000 \\
13 & $13	\times 7	\times 1$ & 20 & 8 & 0.923±0.046 & 0.322±0.066 & 0.902±0.067 & 0.27±0.065 & 0.915±0.048 & 0.733 & 0.967 \\
14 & $14	\times 7	\times 1$ & 21 & 8 & 0.934±0.047 & 0.315±0.078 & 0.906±0.068 & 0.265±0.068 & 0.923±0.049 & 0.733 & 0.967 \\
15 & $15	\times 8	\times 1$ & 23 & 9 & 0.942±0.041 & 0.298±0.064 & 0.913±0.063 & 0.256±0.068 & 0.93±0.046 & 0.733 & 0.967 \\
16 & $16	\times 8	\times 1$ & 24 & 9 & 0.971±0.03 & 0.251±0.05 & 0.955±0.046 & 0.197±0.049 & 0.965±0.039 & 0.800 & 0.967 \\
17 & $17	\times 9	\times 1$ & 26 & 10 & 0.972±0.028 & 0.25±0.055 & 0.953±0.048 & 0.191±0.05 & 0.965±0.038 & 0.800 & 0.967 \\
18 & $18	\times 9	\times 1$ & 27 & 10 & 0.972±0.029 & 0.247±0.047 & 0.952±0.052 & 0.192±0.05 & 0.964±0.039 & 0.800 & 0.967 \\
\end{tabular}
\end{ruledtabular}
\caption{GME detection via ANN accuracy values for total
features from $N=1$ to $N=18$. The ANN network for each $N$ value 
is given by $N
\times {\displaystyle \frac{N + N {\rm mod} \,\,2}{2}} \times 1$; $P_{2}$ and $P_{3}$ refer to total
learning parameters(i.e., weights and biases) for hidden and output
layer,respectively, $\mu(A_{v}) \pm \sigma(A_{v}) $ refer to average validation
set accuracy, $\mu(L_{v}) \pm \sigma(L_{v}) $ to average validation set loss,
$\mu(A_{t}) \pm \sigma(A_{t})$ refer to average test set accuracy and
$\mu(L_{t}) \pm \sigma(L_{t})$ to average test set loss, with standard error
values calculated over 100 validation-test set combinations. $A_{T} \pm
\sigma(A_{T})$ refers to overall accuracy over all experimental states.
$A_{1}$ and $A_{2}$
refer to accuracies obtained via SVM and KNN, respectively.} 
\label{gme_ann_table}
\end{table*}
\subsection{SLOCC Classification via ANN}
\label{classify}
For the training dataset, we generated and validated tripartite states using the
classification parameters specified in Table~\ref{table_1}.
The generation and certification of states was carried out using various Python
3.0 \href{https://www.python.org/} libraries and packages (including Numpy, Scipy,
Sympy), Qiskit\cite{qiskit-2023}, and QuTip\cite{joh-els-2012}, and all these
generated states were included in the training set.

We generated a total of $12 \times 10^{3}$ states with $2 \times 10^{3}$ states
representing each of the six SLOCC inequivalent classes. In this way, the
training set is balanced in terms of distribution of all the SLOCC classes; this
is done so as to fit the SLOCC classification problem and the experimental
dataset which consists 5 states from each of the six SLOCC classes. To
validate and test the ANN model, we experimentally prepared 30 three-qubit
states (choosing 5 states from each SLOCC class), on an NMR quantum processor.  

The training process is represented by trends in the values of the indicators:
training loss, training accuracy, validation loss, and validation accuracy with
respect to the epochs (total training time). These indicators can be used to
tune hyperparameters such as the number of neurons, number of hidden layers,
learning rate, and batch size.  The values of these indicators reflect the
optimization process, which is performed using the stochastic gradient descent
technique called ``Adam'' with a fixed learning rate\cite{sun-ieee-2020}. The
goal of the optimization is to minimize the loss values over the training set
(numerically generated states) and validation set (18 state experimental set)
simultaneously, throughout the epochs.  During optimization, the weights and
biases of each neuron in the hidden layers are updated at each step of the
epoch.  After training, the model with the best weights (i.e., those
corresponding to maximum validation set accuracy and minimum validation loss
over the epochs) is selected. This model is then tested on the test dataset (12
state experimental set) for the particular training instance.

We employed the TensorFlow model Keras library~\cite{abadi-tf-2015} for the
construction, optimization, and analysis of both the SLOCC and GME classification
ANN models. Our approach involved the development of 18
distinct sequential neural network models with the $N^{th}$ ANN model including the
features $1,2, \dots N$ where $N  \in \{1,2,3, \dots 18 \}$. These features were
selected based on their ANOVA F-scores and were organized in descending order of
importance. The input layer consisted of $N$ neurons, while the hidden layer
comprised $6$ neurons and the output layer consisted of $6$ neurons.

The ANN architecture is described in the network column of
Table~\ref{slocc_ann_table} in the form $L_{1} \times L_{2} \times L_{3} $ where
$L_{1}$ refers to the input layer, $L_{2}$ to the hidden layer and $L_{3}$ to
the output layer (the symbol `$\times$' denotes fully connected neurons between
the layers). The ANN parameters are optimized through information processing in
the hidden layers via a process known as forward and backward
propagation~\cite{srivastava-jmach-2014}.  The ``Adam'' optimization process is
driven by a first-order gradient-based stochastic optimization of weights and
biases~\cite{sun-ieee-2020}. 
For each of the 18 ANN models, we randomly sampled 100 
validation:test set combinations for the training process. 
The initial learning rate
was chosen to be $0.001$. The hidden layer and output layer activation functions
are ``linear'' and ``softmax'', respectively. The softmax
function is given by:
\begin{equation}
    \sigma(q_i) = \frac{e^{q_{i}}}{\displaystyle \sum_{j=1}^K e^{q_{j}}}, \textrm{~for~} i=1,2,\dots,K
    \label{eq_softmax}
\end{equation}
where $q_{i}$ = is the value passed from the hidden layers,  $\sigma(q_{i}) = $ the probability of the state belonging to the $i^{th}$ class and $K = $ total number of classes. The stochastic
optimization loss function is ``categorical cross-entropy'' 
which is a standard loss function for multi-class problems 
such as SLOCC classification given by: 
\begin{equation}
    \text{Loss}_{1} = -\frac{1}{M}\sum_{j=1}^{M}
\left(\sum_{i=1}^{K}Y_{i}^{(j)}\log\left[\sigma({X_{i}^{(j)}})\right] \right)
\end{equation}
where M = Total number of states or data points, $K = $ total number of classes,
$Y_{i}^{(j)} = H_{i}(\rho)^{(j)}$ refers to the $i^{th}$ class label value for
the $j^{th}$ state and $\sigma(X_{i}^{(j)}) = \sigma(q_{i}) $ is the predicted
probability for the $i^{th}$ class and $j^{th}$ state. When the
improvement/epoch parameter ``min-delta'' is less than 0.01 for at least 20
epochs(known as ``patience'' parameter), an early stopping monitor is utilized
to end the training process.  The model parameters are then restored to the
values corresponding to best weights and biases ( i.e., particular epoch with
the highest accuracy and lowest loss value). For all the 100 instances of
training and testing over different validation and test set combinations, we
kept the model training fixed at 100 epochs and 1000 batch size. These values were
obtained via trial and error, with the goal of minimizing model training time.
Each epoch of training consists of 120 iterations with each iteration consisting
of 1000 states or data points.  After each epoch the entire training data is
shuffled so as to not bias the learning towards one set of optimal weights and
biases.

\begin{figure}[h]
\centering
\includegraphics[scale=1.0]{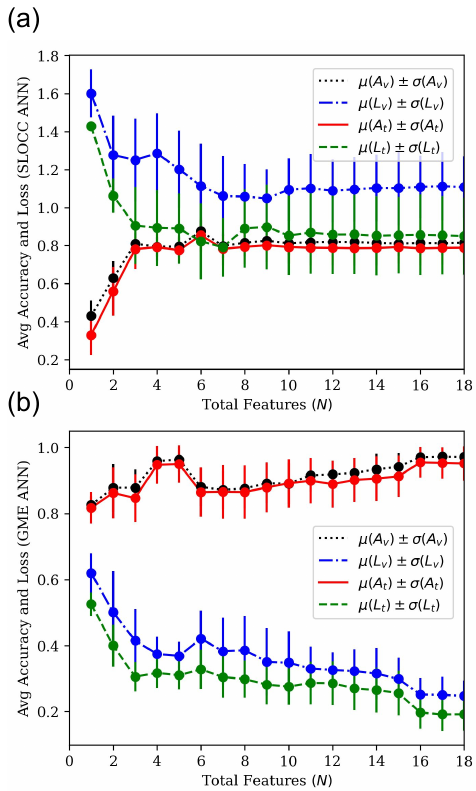}
\caption{(Color online) 
Average accuracy, loss and standard error values for SLOCC 
classification and GME detection via
ANN over total features $N=1$ to $N=18$. (a) SLOCC ANN and (b) GME ANN.
}
\label{fig3-class}
\end{figure}

Figure~\ref{fig3-class} depicts trends of accuracy and
loss over the series of 18 features (in decreasing order of their ANOVA
F-Scores). $\mu({A_{t}})$ and $\mu({A_{v}})$  follow almost the same trend with
a positive correlation of $0.012$ while for $\mu({L_{t}})$  and $\mu({L_{v}})$
the correlation is $0.017$. The sharp increase in $\mu({A_{t}})$  and decrease
in $\mu({L_{t}})$  from $N=1$ to $N=4$ indicated that the features (i.e., density
matrix elements) with the highest F-scores could allow the ANN to classify the
unknown experimental states with sufficiently high accuracy. As mentioned in
Table \ref{slocc_ann_table}, we obtained the highest average test set accuracy
of $0.854 \pm 0.088$ and loss of $0.822 \pm 0.2$  for the 6 features :
$\{\mathbf{Re}(\rho_{07}), \mathbf{Re}(\rho_{05}), \mathbf{Re}(\rho_{06}),
\mathbf{Re}(\rho_{56}), \mathbf{Re}(\rho_{55}), \mathbf{Re}(\rho_{00})\}$ with
an overall accuracy of $0.867 \pm 0.09$. The corresponding ANN model is $6
\times 6 \times 6$ with a total of 24 learning parameters (i.e., weights and
biases) and 18 neurons. The high error values could be attributed to the
stochastic nature of the ANN learning process as well as the problem complexity
(multiclass problem with 6 target variables). 

\subsection{GME Classification via ANN}
\label{gme}
The training dataset for the GME dataset is constructed using the
same method as for the SLOCC problem with the parameters specified 
from Table~\ref{table_1}. The training dataset consists of $12 \times
10^{3}$ states with $2 \times 10^{3}$ from each SLOCC class i.e., $8 \times
10^{3}$ Non-GME states with $4 \times 10^{3}$ GME states. This class
distribution is unbalanced and biased towards the Non-GME class so as to adapt to
the distribution in the 30 state experimental data (i.e, 10 GME and 20 Non-GME
states). As before, we randomly sample 100 validation and test set
combinations consisting of 18 and 12 states, respectively, with the test set kept
unknown to the ANN for each training instance out of 100. We developed 18 ANN
models of the sequence type and dense layers, tailored to correspond to the 18
features derived from manual feature reduction and univariate feature selection.
In this configuration, the input layer was designed to accommodate a variable
number of features, denoted as $N \in \{1, 2, 3, \dots , 17, 18\}$. The number
of neurons in the hidden layer $L_{2}$ was determined using the formula
$\frac{(N + N {\rm mod} \,\,2)}{2}$, and this layer utilized a linear activation function
$\sigma(p) = p$, where $p$ represents the value from the previous layer. The
output layer $L_{3}$ consisted of a single neuron with a sigmoid activation
function:
\begin{equation}
    \sigma(q) = \frac{1} {1 + e^{-q}}
    \label{activation_sigmoid}
\end{equation}
where $\sigma(q)$ denotes the prediction probability of the state belonging to
the GME class. If $ \sigma(q) < 0.5$, it is labeled `0' (Non-GME) and if the
output is $ \sigma(q) \geq 0.5$, it is labeled `1' (GME). The loss function used
is `binary crossentropy' given by: 
\begin{equation}
    \text{Loss}_{2}=\frac{-1}{M}
\sum_{j=1}^{M}{\left[Y^{(j)}\log(X^{(j)}) + (1 - Y^{(j)})\log(1 -
X^{(j)})\right]}
    \label{loss_binarycrossentropy}
\end{equation}
where $Y^{(j)}$ is the actual label or value of the target variable for the
$j^{th}$ state in the set of $M$ states (i.e.,
$\mathrm{B(\rho)}^{(j)}$) and
$X^{(j)} = \sigma(q^{(j)})$.  For all the 18 ANN models, the learning rate,
min-delta and patience parameters are kept at $0.001, 0.01$ and $20$,
respectively. With the aim of minimizing training time, we arrived at the $100$
epochs and $500$ batch size via trial and error. Each epoch consists of $120$
iterations with each iteration consisting of 100 states from the training
dataset. After each epoch the entire training data of $12000$ was shuffled. 

Figure~\ref{fig3-class} (b) shows the obtained trends of average accuracy and
loss values over the validation and test set for the series of 18 features
arranged in decreasing order of their ANOVA F-score. The correlation values for
the test set average accuracy $\mu({A_{t}})$ and validation set average accuracy
$\mu({A_{v}})$ is $0.002$ while for test set average loss $\mu({L_{t}})$ and
validation set average loss $\mu({L_{v}})$ is $0.007$. Again, there is an
increase in $\mu({A_{t}})$ and decrease in $\mu({L_{t}})$ with $N = 1$ from $N =
5$ which indicates that only a few features with higher ANOVA F-scores can be used
for the GME classification of unknown test set states. Table~\ref{gme_ann_table}
shows the average accuracies and errors, the smallest ANN model with the highest
test set accuracy of and loss of is given by $4 \times 2 \times 1$ with the 4
features being $( \mathbf{Re}(\rho_{07}), \mathbf{Re}(\rho_{56}),
\mathbf{Re}(\rho_{05}), \mathbf{Re}(\rho_{06}))$. The overall accuracy over 30
states is $0.954 \pm 0.046$. The model consists of a total of 9 learning
parameters (weights and biases) and 7 neurons. The standard error in accuracies
is relatively low compared to the case of SLOCC ANN as now the problem is a mere
binary classification problem with only 1 target variable. 

\begin{figure}[h] 
\centering
\includegraphics[scale=1.0]{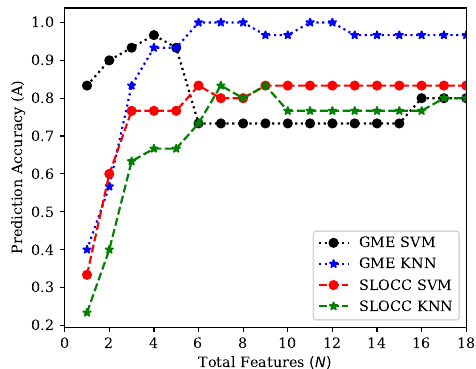} 
\caption{(Color online) Accuracy of SVM and KNN models over the total features
$N=1$ to $N=18$.} 
\label{fig4-gme} 
\end{figure} 
\section{Benchmarking ANN Performance} 
\label{benchmark}
In this section, we compare the performance of the optimal ANN
models with previously used ML techniques in entanglement detection and
classification such as SVM (Support Vector Machines) and KNN (K-Nearest
Neighbors) as well as with entanglement classifiers,
namely, correlation tensors and 3-tangle for SLOCC classification and correlation
tensors for GME detection. 
\subsection{Comparison with SVM and KNN} 
\label{svm} 
SVM and KNN algorithms have
been recently used to characterize entanglement in
two-qubit~\cite{hiesmayr-nsr-2021} and four-qubit
systems~\cite{vintskevich-pra-2023}.  We evaluated the performance
of these models by training them on the same training dataset of $12 \times
10^{3}$ which was used for GME and SLOCC ANN. These models do not require a
validation set hence, these are tested over the complete 30 state experimental
dataset which was kept unknown to both the SVM and KNN models.
For the SLOCC inequivalent entanglement classification problem, the class labels
are in the integer-encoding form instead of the one-hot encoding form.

We used Python Sklearn library for implementing SVM and KNN models, for both
GME and SLOCC classification of the experimental dataset. We considered a
linear SVC kernel with a $0.1$ regularization (a parameter which allows one to
set the tolerance threshold for misclassification).  For GME, we considered a
KNN model $k = 18$ nearest neighbors, ``uniform'' weights and $p=1$ (i.e.,
Manhattan distance $\sum^n_{i=1}|X_i-Y_i|$ with $X_i$ and $Y_i$ being $d \times
1$ dimensional vector) as the Minkowski (distance) metric. For SLOCC, KNN
parameters are $k=20$, ``distance'' weights (i.e., closer points have higher
weightage in deciding label of unknown state) and $p=2$ (i.e., Euclidean
distance $[\sum^n_i(X_i -Y_i)^{2}]^{\frac{1}{2}}$ ) as the Minkowski metric.

Figure~\ref{fig4-gme} shows trends of SVM and KNN accuracies for the GME and
SLOCC classification problems, with total number of features $N$. For the case
of GME, both SVM and KNN accuracies increase till $N=4$ after which only KNN
accuracy is able to reach $1.00$ while SVM accuracy declines and only slightly
increases after $N=16$. For SLOCC, the situation is different: $N = 6$
represents the lowest number of total features where the SVM's accuracy becomes
comparable to $N > 9$ (0.833). In contrast, KNN reaches this accuracy threshold
at $N = 7$.

Table~\ref{slocc_ann_table} contains the accuracies obtained via SVM and KNN
for the SLOCC classification problem as, $A_{1}$ and $A_{2}$.For the case of
SLOCC, for $N > 10$ the SVM performs slightly better than ANN but KNN lags
behind at $N=6$, $\mu(A_{t}) = 0.854$ while $A_{1} = 0.833 $ and $A_{2} = 0.733
$. For $N < 6$, the SLOCC ANN model performs slightly better than both SVM and
KNN models. For the case of GME in the value range of $ N = 6, 7, 8$ the KNN
model outperforms ANN with $A_{2} = 1.00$ but at low feature number of $N=4$,
the ANN model accuracy is at $\mu(A_{t}) = 0.954$ while  for SVM and KNN it is,
$A_{1} = 0.967 $ and $A_{2} = 0.933$, respectively. For extremely low feature
range of $N = 2, 3$ SVM performs slightly better than GME ANN with accuracies
$0.9$ and $0.933$, respectively. Table~\ref{table_8} shows the comparison of
SVM and KNN the optimal SLOCC and GME ANNs, being, $6 \times 6 \times 6$ and $4
\times 2 \times 1$, respectively. In the context of GME classification, we
observe varying model performance across different feature numbers.
Specifically, for N values of 6, 7, and 8, the KNN model demonstrates higher
performance compared to the ANN with an accuracy score $A_{2}$ of $1.00$.
However, when the total feature number is reduced to $N = 4$, the ANN model
achieves a commendable accuracy of $\mu(A_{t}) = 0.954$. In contrast, SVM and
KNN models achieve accuracy scores of $ A_{1} = 0.967$ and $A_{2} = 0.933$,
respectively. For an extremely reduced feature dimension of $N = 2$ and $3$,
SVM exhibits a slightly better performance when compared to the GME ANN, with
accuracy scores of $0.9$ and $0.933$, respectively. A comprehensive comparison
of SVM and KNN models with the optimal SLOCC and GME ANN configurations is
provided in Table~\ref{table_8}, with the respective ANN models as 6 × 6 × 6
for SLOCC and 4 × 2 × 1 for GME. Since we are comparing the performance over
unknown datasets, we have used $\mu(A_{t})$ for comparison.

\begin{table}[h]
\begin{tabular}{P{3cm}P{1.7cm}P{1.5cm}P{1.5cm}}
\hline \hline 
Average Test Set Accuracy & 
 ANN $\mu(A_{t})$ & SVM $A_{1}$ & KNN $A_{2}$ \T\B \\
\hline 
{\rm SLOCC $(N = 6)$}   & 0.854 & 0.833 & 0.733 \T\\
{\rm GME $(N =4)$}  & 0.948 & 0.967 & 0.933  \B\\
\hline 
\end{tabular}
\caption{Comparison of the accuracy of ANN ($\mu(A_{t})$),
SVM ($A_{1}$) and KNN ($A_{2}$) models for
SLOCC and GME entanglement classification.}
\label{table_8}
\end{table}

\begin{table}[ht]
\begin{tabular}{P{2.5cm}P{2.0cm}P{2.8cm}}
\hline \hline 
\T Average Test Set Accuracy &  ANN $\mu(A_{t})$  & Correlation Tensors \& 3-tangle  \T\B\\
\hline 
$\mathrm{SLOCC} (N = 6)$  &$0.854$  & $0.80$ \T\\
$\mathrm{GME} (N = 4)$ & $0.948$   &  $1.00$ \B\\
\hline
\end{tabular}
\caption{Comparison of ANN results with total number of features, 
$N=4$ for GME and $N=6$ for SLOCC, with those obtained using 
3-tangle $\tau_{123}$ and correlation tensors rank
$\{R(T_{\underline{1}23}), R(T_{\underline{2}13}), R(T_{\underline{3}12}\}$,
respectively.}
\label{table_9} 
\end{table}

\begin{table*}
\centering
\begin{tabular}{|P{2.3cm}|P{0.9cm}c|P{0.9cm}c|P{0.9cm}c|P{1.2cm}P{0.9cm}|P{0.9cm}cc|}
           \hline
			\textbf{Obs.} $\boldsymbol{\rightarrow}$ 
			& \multicolumn{2}{P{2.1cm}|}     
                {\textbf{R(T}$\boldsymbol{_{\underline{1}23}}$)}
                & \multicolumn{2}{P{2.1cm}|} {\textbf{R(T}$\boldsymbol{_{\underline{2}13}}$)}
                & \multicolumn{2}{P{2.1cm}|}{\textbf{R(T}$\boldsymbol{_{\underline{3}12}}$)}& \multicolumn{2}{P{2.4cm}|}{\textbf{3-tangle}      ($\boldsymbol{\tau_{123}}$)} 
			& \multicolumn{3}{P{2.6cm}|}{\textbf{GME-R(T}$\boldsymbol{_{\underline{i}jk})}$} \T\B\\
			\hline
			\textbf{State} $\boldsymbol{(F) \downarrow}$ &
                \textbf{Th.} & \textbf{Ex.} & \textbf{Th.} & \textbf{Ex.} & \textbf{Th.} & \textbf{Ex.} & 
                \textbf{Th.} & \textbf{Ex.} &
                \textbf{Th.} & \textbf{Ex.} & \textbf{Ac.} \T\B \\
			\hline 
   
			$\mathrm{SEP_{1}}(0.9755)$ 
			& 1	& 1	& 1  & 	1 &	1 &	1 
                & 0 & 0 
			  & 0 & 0 &1  \T\\
   
			$\mathrm{SEP_{2}}(0.9585)$ 
			& 1 &	1 &	1 &	1 &	1 &	1 
                &	0&	0
			  & 0 & 0 & 1  \T\\
   
			$\mathrm{SEP_{3}}(0.9329)$ 
			& 1	 & 1 & 	1 & 1 & 	1 &	1	
                & 0 & 0 
			  & 0 & 0 &1 \T\\
   
			$\mathrm{SEP_{4}}(0.9482)$ 
			& 1 & 	1 &	 1 & 1 & 1	& 1 
                & 0 & 0 
			  & 0 & 0 &1 \T\\
   
			${\mathrm{SEP_{5}}(0.9848)}$ 
			& 1 & 1 &	1 & 	3 & 1 &	3  	
                & 0 &	0 
			  & 0 & 0 &1 \T\B\\
			\hline
   
			${\mathrm{BS1_{1}}(0.9705)}$ 
			& 1	& 1&	3&	2&	3&	2
                & 0 &	0
			  & 0 & 0 &1 \T\\

			$\mathrm{BS1_{2}}(0.9807)$
			& 1&	1&	3&	1&	3&	1	
                & 0 &	0 
			  & 0 & 0 &1 \T\\

			${\mathrm{BS1_{3}}(0.9381)}$ 
			& 1&	3&	3&	1&	3&	3	
                & 0 &	0
			  & 0 & 0 &1  \T\\

			${\mathrm{BS1_{4}}(0.9731)}$ 
			& 1&	3&	3&	1&	3&	3	
                & 0 &	0
			  & 0 & 0 &1 \T\\

			${\mathrm{BS1_{5}}(0.9447)}$ 
			& 1&	3&	3&	1&	3&	3	
                & 0 &	0
			  &0 & 0 &1 \T\B\\
			\hline

			${\mathrm{BS2_{1}}(0.9687)}$ 
			& 3&	3&	1&	1&	3&	3	
                & {0} &	{0}
			  & {0} & {0} &1 \T\\

			${\mathrm{BS2_{2}}(0.9538)}$ 
			& 3&	3&	1&	1&	3&	3 
                &	0 &	0
			  & {0} & {0} &1  \T\\

			$\mathrm{BS2_{3}}(0.9514)$ 
			& 3&	3&	1&	1&	3& 3	
                &	0 &	0.1
			  & 0 & 0 &1 \T\\

			$\mathrm{BS2_{4}}(0.9262)$ 
			& 3&	3&	1&	1&	3&	3	
                &	0 &	0
			  & 0 & 0 &1 \T\\

			$\mathrm{BS2_{5}}(0.9718)$ 
			& 3&	3&	1&	3 &	3	& 1 
                &	0&	0
			  & 0 & 0 &1 \T\B\\
                \hline

			$\mathrm{BS3_{1}}(0.9807)$ 
			& 3&	3&	3&	3&	1&	1  
                &	0&	0
			  & 0 & 0 &1 \T\\

			$\mathrm{BS3_{2}}(0.9598)$ 
			& 3&	3&	3&	3&	1&	1  
                &	0&	0
			  & 0 & 0 &1 \T\\

			$\mathrm{BS3_{3}}(0.9635)$ 
			& 3&	3&	3&	3&	1&	1	
                &	0&	0	 
			  & 0 & 0 &1 \T\\

			$\mathrm{BS3_{4}}(0.9544)$ 
			& 3&	3&	3&	3&	1&	1	
                &	0&	0
			  & 0& 0 &1 \T\\

			$\mathrm{BS3_{5}}(0.9753)$ 
			& 3&	3&	3&	3&	1&	1  
                &	0&	0
			  & 0 & 0 &1 \T\B\\
			
			\hline
   
			$\mathrm{W_{1}}(0.9489)$ 
			& 3&	3&	3&	3&	3&	3	
                &	0&	0
			  & 1 & 1 &1  \T\\

			$\mathrm{W_{2}}(0.8978)$ 
			& 3&	3&	3&	3&	3&	3	
                &	0&	0
			  & 1 &  1 &1  \T \\

			${\mathrm{W_{3}}(0.9669)}$ 
			& 3&	3&	3&	3&	3&	3
                &	0&	0
			  & 1 & 1 &1  \T\\

			$\mathrm{W_{4}}(0.9253)$ 
			& 3&	3&	3&	3&	3&	3
                &	0&	0
			  & 1 & 1 &1 \T \\

			$\mathrm{W_{5}}(0.9628)$
			& 3&	3&	3&	3&	3&	3
                &	0&	0
			  & 1 & 1 &1  \T\B\\
			\hline
   
			$\mathrm{GHZ_{1}}(0.9323)$ 
			&3&	3&	3&	3&	3&	3
                &	  0.6 & 0.3
			  & 1 & 1 &1  \T\\

			$\mathrm{GHZ_{2}}(0.9493)$ 
			& 3&	3&	3&	3&	3&	3
                &	0.9 & 0.6 
			  & 1 & 1&1  \T\\

			$\mathrm{GHZ_{3}}(0.9485)$ 
			& 3&	3&	3&	3&	3&	3
                &	0.8 & 0.5
			  & 1 & 1 &1  \T\\

			$\mathrm{GHZ_{4}}(0.8743)$ 
			& 3&	3&	3&	3&	3&	3
                &	0.5 & 0.4
			  & 1 & 1 &1 \T \\

			$\mathrm{GHZ_{5}}(0.9549)$
			& 3&	3&	3&	3&	3&	3
                &	0.9 & 0.3
			  & 1 & 1 &1  \T\B\\		
     \hline
\end{tabular}
\caption{Calculated values of  3-tangle  $\tau_{123}$ and correlation
tensors ranks $\{R(T_{\underline{1}23}), R(T_{\underline{2}13}),
R(T_{\underline{3}12})\}$ for each of the $30$ experimentally prepared states,
denoted by ``Ex.'' and their theoretical counterparts, denoted by ``Th.''. The
fidelities of the experimentally prepared states are written in the brackets.
``Ac.'' refers to prediction accuracy of each state via 
the correlation tensor rank
method.}
\label{table_10}
\end{table*}

\subsection{Comparison with entanglement measures}
\label{3tangle}
We compared results obtained using the ANN model 
for GME ($N=4$) and for SLOCC ($N=6$) 
entanglement classification
with those
obtained by using entanglement classification methods such as 
3-tangle $\tau_{123}$, and 
ranks of correlation tensors
$\{R(T_{\underline{i}jk})\}$.  The accuracy for each method was calculated
as the ratio of correctly predicted labels to the total number of labels. We
used Numpy's linalg library (which uses Singular Value Decomposition (SVD)) to
calculate the correlation matrix ranks. The SVD tolerance for the experimental
correlation tensors was kept at ~$2.5 \times 10^{-1}$, whereas it was kept at
~$1 \times 10^{-10}$ for the reference theoretical states.

For SLOCC classification via correlation tensor ranks and
3-tangle, the accuracy for each class was obtained as: (SEP = $\frac{4}{5}$=
0.8,  BS${1}$ = $\frac{1}{5}$ = 0.2, BS${2}$ = $\frac{4}{5}$ =0.8, BS${3}$ =
$\frac{5}{5}$= 1.0, W = $\frac{5}{5}$ = 1.0, GHZ = $\frac{5}{5}$ =1.0),
implying an overall SLOCC accuracy of $\frac{24}{30}$ = 0.8 = $80\%$. Since a
rank greater than 1 (2 or 3) implies the presence of entanglement, the first
state of BS1 is considered correctly classified. For the case of GME/Non-GME
classification, the correlation tensors have an accuracy of $1.00 = 100\%$.  In
comparison with GME ANN, the correlation tensor method outperforms with
accuracy of $1.00$ but it requires the calculation of $13$ expectation values,
although GME $(N=4)$ ANN classifies the unknown states with an accuracy of only
$0.948$, it does so with only 4 real density matrix elements. For the case of
SLOCC, $(N=6)$ ANN outperforms the local entropy and 3-tangle method with an
accuracy of $0.854$.  These results are summarized in Table~\ref{table_9}.

Table~\ref{table_10} contains the obtained values of 3-tangle
and correlation tensor ranks for the $30$ experimentally prepared NMR states
(denoted by ``Ex.'') as well as their theoretical counterparts (denoted by
``Th.'').  The first column consists of the class label with the state fidelity
(F),  and the subscript denoting the different states of the same class.  The
``Ac.'' subcolumns indicate whether the predicted label for each state Ac. $\in
\{0,1\}$ is correct (0) or incorrect (1).  The
``GME-Rank(T)$_{\underline{i}jk})$'' column presents the GME classification
values determined using the correlation tensor rank method for both theoretical
and experimental states. A value of `1' represents GME, while `0' corresponds to
``NON-GME''. 
\section{Conclusions}
\label{concl}
We demonstrate the efficacy of an ANN model in correctly identifying a
three-qubit state from one of the six SLOCC inequivalent entanglement classes
and in detecting the presence of GME in the state.  We utilize the generic form
of the tripartite states to reduce the number of relevant density matrix
elements to 18 (14 real and 4 imaginary), thereby reducing the dimensionality
of the problem.  The ANN models are trained on numerically generated
three-qubit quantum states, and validated and tested on a $30$ state
experimental dataset generated on a three-qubit NMR quantum processor.  We
demonstrate that it is possible to obtain high accuracies for GME detection and
identification of the SLOCC class via the ANN model, with total number of
features as low as $4$ and $6$, respectively.  

We compare the performance of the trained ANN models with alternative ML
classification schemes such as SVM and KNN as well as entanglement measures
3-tangle and correlation tensors.  The ANN model with 6 features performed with
higher accuracy in identifying the SLOCC-inequivalent entanglement classes as
compared to SVM, KNN and correlation tensor and 3-tangle methods, while for the
case of GME, the ANN model with 4 features performed at par with SVM, KNN and
the correlation tensors method.  The relatively higher standard error of SLOCC
ANN as compared to GME ANN can be attributed to the difference in problem
complexities, with the SLOCC classification being a multi-class problem while
GME detection is a binary classification problem.  Our results demonstrate that
ANNs are promising alternatives to experimentally demanding methods such as
full quantum state tomography and witness-based methods in characterizing
entanglement in noisy quantum states.

\vspace*{1cm}
\begin{acknowledgments}
The NMR experimental data were obtained on a Bruker Avance-III 600 MHz FT-NMR
spectrometer at the NMR Research Facility at IISER Mohali.  
\end{acknowledgments}

%
\end{document}